\documentclass[%
aip,
amsmath,amssymb,
reprint,%
]{revtex4-2}

\usepackage{graphicx}% Include figure files
\usepackage{dcolumn}% Align table columns on decimal point
\usepackage{bm}% bold math
\usepackage[utf8]{inputenc}
\usepackage[T1]{fontenc}
\usepackage{mathptmx}
\usepackage{etoolbox}
\usepackage{textcomp}
\usepackage{latexsym}

\usepackage[dvipsnames]{xcolor}
\usepackage[normalem]{ulem}

\makeatletter
\def\@email#1#2{%
	\endgroup
	\patchcmd{\titleblock@produce}
	{\frontmatter@RRAPformat}
	{\frontmatter@RRAPformat{\produce@RRAP{*#1\href{mailto:#2}{#2}}}\frontmatter@RRAPformat}
	{}{}
}%

\makeatother
\renewcommand{\selectlanguage}[1]{}
\begin{document}
	
	\preprint{AIP/123-QED}
	
	\title[Miniaturized vacuum package for magneto-optical trapping of strontium]{Miniaturized vacuum package for magneto-optical trapping of strontium}
	% Force line breaks with \\
	\author{J. Pick}
	\email{julian.pick@dlr.de}
	\affiliation{Deutsches Zentrum für Luft- und Raumfahrt e.V., Institut für Satellitengeodäsie und Inertialsensorik, Callinstraße 30b, 30167 Hannover, Germany}
	\author{E. Henker}
	\author{F. Löwinger}
	\author{R. Vogt}
	\author{S. Hüttl}
	\author{A. Trützschler}
	\affiliation{VACOM Vakuum Komponenten \& Messtechnik GmbH, In den Brückenäckern 3, 07751 Großlöbichau, Germany}
	\author{J. Voß}
	\author{S. Hirt}
	\author{M. Schulz-Ruhtenberg}
	\affiliation{LPKF Laser \& Electronics SE, Osteriede 7, 30827 Garbsen, Germany}
	\author{A. Buchta}
	\author{A. Kassner}
	\author{F. Dencker}
	\author{M. C. Wurz}
	\affiliation{Leibniz Universität Hannover, Institut für Mikroproduktionstechnik, An der Universität 2, 30823 Garbsen, Germany}
	\author{S. Hannig}
	\author{S. Callegari}
	\author{S. A. Bondza}
	\affiliation{Physikalisch-Technische Bundesanstalt, Bundesallee 100, 38116 Braunschweig, Germany}
	\author{J. Kruse}
	\affiliation{Deutsches Zentrum für Luft- und Raumfahrt e.V., Institut für Satellitengeodäsie und Inertialsensorik, Callinstraße 30b, 30167 Hannover, Germany}
	\author{T. Leopold}
	\affiliation{Deutsches Zentrum für Luft- und Raumfahrt e.V., Institut für Satellitengeodäsie und Inertialsensorik, Callinstraße 30b, 30167 Hannover, Germany}
	\affiliation{LPKF Laser \& Electronics SE, Osteriede 7, 30827 Garbsen, Germany}
	\author{R. Schwarz}
	\author{C. Klempt}
	\affiliation{Deutsches Zentrum für Luft- und Raumfahrt e.V., Institut für Satellitengeodäsie und Inertialsensorik, Callinstraße 30b, 30167 Hannover, Germany}

	\date{\today}
	
	\begin{abstract}
	Quantum sensors, like optical lattice clocks, undergo a continuous development from lab-based towards mobile systems. A key aspect in the miniaturization of the experimental setups is the development of compact atom sources. Conventional atom sources for alkaline-earth-like elements consist of a high-power oven and a six-beam magneto-optical trap (MOT) inside of vacuum chamber with extensive flanges, viewports and electrical feedthroughs. Here we present a highly compact vacuum package utilizing key technologies for future quantum sensor systems: A chip-based low-power atomic oven, a planar grating MOT chip, and a miniaturized vacuum pump all held inside an additively manufactured titanium vacuum chamber, featuring custom-sized vacuum flanges. In this miniaturized setup, spanning a volume of only $750\,$ml, we trap up to $10^5$ Sr atoms in a MOT, requiring an oven heating power below $1\,$W.
	\end{abstract}
	
	\maketitle

	\section{\label{sec:introduction}Introduction}
	Alkaline-earth-like atoms, most prominently strontium and ytterbium, are key species in quantum simulation and quantum sensing, most notably in optical lattice clocks\cite{takamoto_optical_2005, ludlow_optical_2015}. As their development progresses, quantum sensors are becoming more efficient in size, weight and power consumption (SWaP), while simultaneously improving in reliability and robustness. Being at the verge of transitioning from large experiments in fundamental science to field-deployable instruments \cite{origlia_towards_2018, ohmae_transportable_2021, bothwell_deployment_2025, nosske_transportable_2025}, further development in improving the SWaP budgets is ongoing.
	
	A significant part of the SWaP budget of conventional quantum sensors stems from the atom source. For alkaline-earth-like elements, the atom preparation typically relies on a high-power atomic oven and a two-colour six-beam magneto-optical trap (MOT), requiring a large and complex optical setup. The complexity of a MOT can be reduced by employing in-vacuum optical elements, reflecting or diffracting a single input beam into the beam geometry required for three-dimensional trapping and cooling. Examples include pyramid reflectors \cite{lee_single_1996, bowden_pyramid_2019, pick_compact_2024}, tetrahedral reflectors \cite{vangeleyn_single_2009}, planar diffraction gratings \cite{vangeleyn_laser_2010, nshii_surface_2013, sitaram_confinement_2020, bondza_two_2022, barker_grating_2023}, or a Fresnel reflector \cite{bondza_achromatic_2024, pick_compact_2024}.
	For application in an optical lattice clock, the atomic oven must provide temperatures of up to $500\,^\circ$C \cite{nosske_transportable_2025}, in order to evaporate a sufficient flux of atoms. This typically leads to a high power consumption as well as large experimental setups. Progress towards more power-efficient heating of atomic ovens has been made by employing in-vacuum heating \cite{schioppo_compact_2012,Fartmann_Ramsey_2025}. A significant further reduction of the required electrical heating power and the system size was achieved using chip-based atomic ovens, based on silicon \cite{schwindt_highly_2016,Kumar_Fast_2025b}, or fused silica \cite{pick_low_2025}.
	
	The vacuum chamber represents another limiting factor for miniaturization. Besides the chamber itself, optical viewports, electrical feedthroughs and standardized vacuum flanges often dictate the overall dimensions of the system. Commercial ConFlat (CF) components are generally available only in standardized sizes, often resulting in vacuum systems that are larger than functionally required. In addition, chamber designs must satisfy practical manufacturing constraints, further restricting the geometric flexibility. Recently, additive manufacturing of an aluminum vacuum chamber has been reported \cite{cooper_additively_2021}, opening the path towards chamber designs that are not possible with conventional subtractive machining methods. This includes weight-optimized surface structures that preserve mechanical stability, integrated water cooling channels, and complex multi-chamber designs tailored to specific experimental requirements.

	In order to reach the required vacuum pressure, ion getter pumps and non-evaporable getter pumps are commonly employed. Traditional ion getter pumps rely on strong magnetic fields that trap electrons in spiral paths, increasing the chance of ionizing residual gas molecules and implanting them into a titanium getter cathode. The magnets and massive flanges, however, make such pumps unsuitable for compact or mobile quantum devices, and the magnetic fields can interfere with sensitive experiments.
    To deal with these limitations, magnet-free designs have been proposed that use a field-emission electron source, specifically field emitter arrays (FEAs) \cite{Basu_Miniature_2015, Basu_Nanostructured_2015, Basu_electrostatic_2016}. In these devices, electrons emitted from a FEA ionize residual gas molecules, whose ions are subsequently captured by a reactive getter electrode. The ion current collected relates directly to chamber pressure, so the device can function both as an active pump and an in-situ pressure gauge.
    Recent developments of diced silicon-based FEA chips integrated into glass-silicon substrates\cite{Basu_Nanostructured_2015, Edler_Silicon_2021, Schels_In_2022, buchta_novel_2023, Hausladen_Measurement_2024} in magnet-free pumps make this approach particularly attractive for chip-scale ultra-high-vacuum systems and compact quantum sensors\cite{Kassner_Miniaturized_2023}.

    Here, we present a highly compact source of trapped and laser-cooled strontium atoms that combines several complementary miniaturization technologies. The central part of the setup consists of a stack of two chips, which are used to evaporate Sr atoms and to trap and cool them in a MOT. Atoms are evaporated from a low-power microstructured oven and then trapped in a MOT that is generated by a grating MOT chip from a single incident laser beam. This chip stack is integrated into an additively manufactured titanium vacuum chamber with custom-sized CF flanges, resulting in a total vacuum package volume of only $750\,$ml. Using an oven heating power below $1\,$W, we trap up to $10^5$ Sr atoms in a grating MOT. In addition, we present a compact magnet-free ion getter pump, that can be integrated directly into the vacuum chamber, further advancing the realization of fully miniaturized quantum sensors.

    \begin{figure*}
		\centering
		\includegraphics[width = 0.85\textwidth]{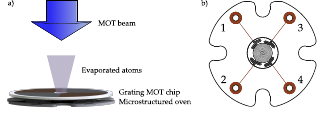}
		\caption{a) Stack of two chips that are used to evaporate, trap and cool Sr atoms. The gray shaded area represents the evaporated atoms from the oven, that can enter the trapping volume through the central hole of the grating MOT chip. b) Sketch of the back side of the microstructured oven. The brown numbered areas are copper pads used for the electrical connection. Gray lines indicate the platinum metallization. Adapted with permission from Ref. \cite{pick_compact_2026}. \label{fig:oven_with_gMOT}}
	\end{figure*}

%########## D E S I G N ##############################
    \section{Design}
    The vacuum package that is used for magneto-optical trapping of Sr consists of different miniaturized components, that are described in detail in the following.
    
    \begin{figure*}
		\centering
		\includegraphics[width = 0.85\textwidth]{"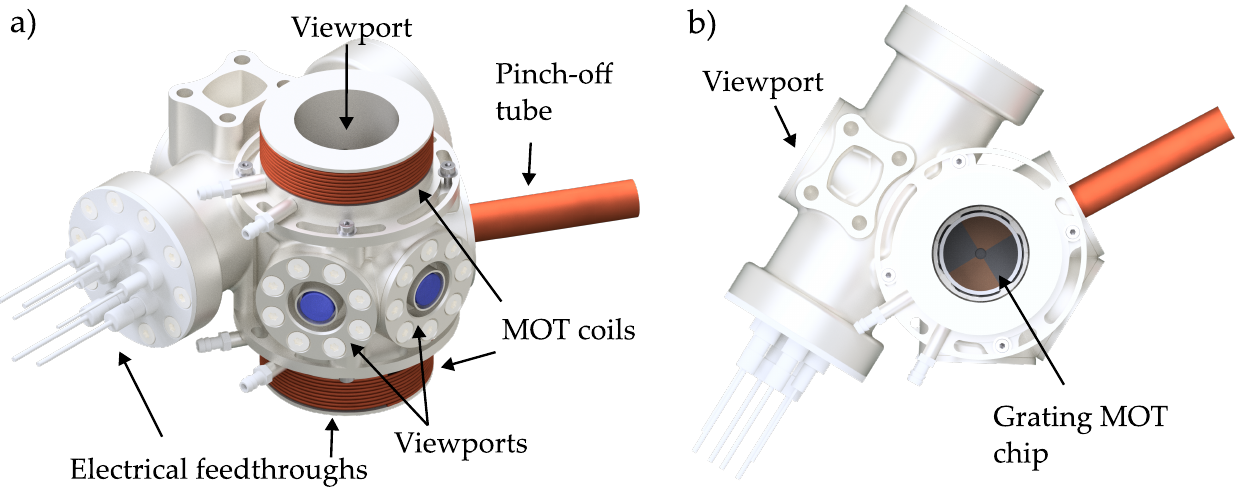"}
		\caption{Rendered drawing of the vacuum chamber. a) View from the side. b) View from the top. The top viewport is not shown, so that the grating MOT chip is visible. Adapted with permission from Ref. \cite{pick_compact_2026}. \label{fig:vacuum_chamber}}
	\end{figure*}
    \subsection{Atomic source}
    Many cold atom experiments are based on high-power effusive ovens that are either self-built \cite{schioppo_compact_2012,wodey_robust_2021} or commercially available \cite{Letellier_Loading_2023, AbdelKarim_Single_2025}, with developments towards miniaturized, chip-based designs \cite{schwindt_highly_2016,Kumar_Fast_2025b}. 
    Here we demonstrate the application of an oven chip based on microstructured fused silica, which is described and characterized in detail in Ref.\cite{pick_low_2025}. 
    
   %####################   O V E N / g M O T   ####################
    
    The central part of the microstructured oven is the atomic reservoir filled with strontium. The back of the reservoir has a platinum metallization in the shape of a Fermat spiral, allowing to electrically heat the reservoir to enable atomic evaporation. Via four narrow mounting springs it is connected to the outer support structure that features four through-glass-vias for the electrical contact. The narrow springs yield a significant thermal insulation from the support structure, enabling low-power operation. The oven has been manufactured by laser-induced deep etching (LIDE). The outer support structure is larger compared to Ref.~\cite{pick_low_2025}, with an outer diameter of $23\,$mm. This makes the oven more convenient to handle and simplifies the electrical connection.
    The microstructured oven is operated in combination with a grating MOT chip for single-beam operation (Fig.~\ref{fig:oven_with_gMOT}). Both components were fabricated from fused silica wavers of $0.5\,$mm thickness. In combination both components form a highly compact platform to generate a strontium MOT based on a dual-chip design.
    
    The grating MOT chip has four segments and an outer diameter of $2\,$cm. It is a rebuild of a chip that has been designed, tested and characterized for two-stage trapping and cooling of strontium\cite{bondza_two_2022}. It requires only a single incident laser beam from the top to generate all the required beams for three-dimensional trapping and cooling. It has a central hole of $3\,$mm diameter that allows the evaporated atoms to enter the trapping volume from the bottom.
    
    This assembly is mounted on a vacuum flange that features four electrical feedthroughs, two of which are soldered to the copper pads of the oven for electrical heating. The other two were not soldered, since it became apparent that a resistance measurement of the heated metal did not give information about the temperature. Also, fixing the oven only at two points reduced the risk of mechanical stress causing damage.

    \subsection{Vacuum chamber}
    The vacuum chamber was designed to house the atomic source and vacuum pumps with a very small overall volume.  The chamber body was additively manufactured by selective laser melting of titanium, and subsequently, knife edges and threads were added by conventional machining. The additive manufacturing is not required by the chamber design, but is used as a proof-of-principle demonstration of the usability of this technique for titanium vacuum chambers. The outgassing of additively manufactured titanium has been investigated using a separate test chamber in the shape of a CF40 cube. Here, an outgassing rate of less than $1\,\%$ compared to a stainless-steel cube has been observed. For this comparison, both parts underwent the same cleaning procedure and were baked at $200\,^\circ$C for $24\,$h.
    
    In order to minimize the chamber size while providing the required clear apertures for all laser beams, custom-sized flanges were manufactured. The four larger ones have an inner diameter of $26\,$mm and the four smaller ones have a $14\,$mm inner diameter. In contrast to the standard CF flange design, they are inserted into the openings of the chamber, minimizing the overall footprint. UHV-compatible viewports were produced by glueing sapphire windows into the flanges. The design is shown in Fig.~\ref{fig:vacuum_chamber}. A copper pinch-off tube connects the vacuum chamber to a turbo pump and to a $2\,\mathrm{l}/\mathrm{s}$ ion getter pump, in order to monitor the vacuum pressure. The magnetic quadrupole field for the MOT is generated by two coils that are wound on internally water-cooled titanium mounts. The coil mounts are directly attached to the vacuum chamber. Vacuum pumps can be placed in the back part of the chamber. A non-evaporable getter pump (SAES St 172) is mounted on an electrical feedthrough flange, so that it can be electrically heated for initial activation. On the opposite side, a blind flange is mounted. In the future, it can be replaced by an electrical feedthrough holding a miniaturized ion getter pump\cite{buchta_novel_2023, buchta_glass_2024}, which was tested inside an identical second vacuum chamber and which is described in the following section.
    The assembled vacuum chamber including the MOT coils and the electrical feedthroughs spans a volume of $750\,$ml. 
    
    \subsection{Vacuum pump}
\begin{figure*}
		\centering
		\includegraphics[width = 0.8\textwidth]{"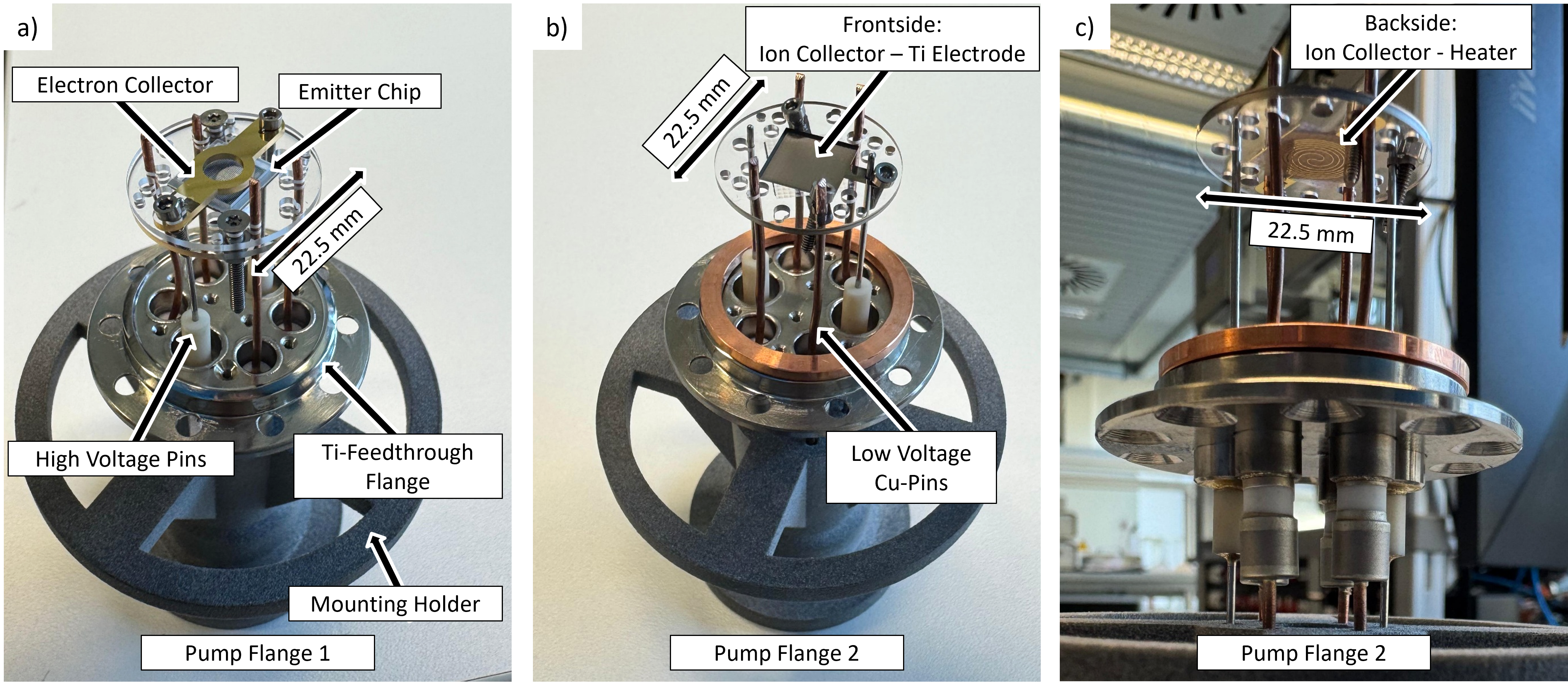"}
		\caption{Miniaturized ion getter pump assembly mounted on two CF titanium feedthrough flanges of the vacuum chamber. Each flange provides two high‑voltage and four low‑voltage current feedthroughs and is attached to a 3D‑printed mounting structure for mechanical stability. The silicon emitter chip with the ring‑shaped electron collector is assembled on pump flange 1 (a), while the ion‑collector assembly is mounted on pump flange 2 (b,c). Image b shows the front side of the titanium‑coated ion‑collector electrode, and image c the back side with the integrated heater. \label{fig:pump_assembly}}
	\end{figure*}

    Because several high‑voltage connections are required for pump operation, the ion getter pump had to be distributed over two CF feedthrough flanges. This is shown in Fig.~\ref{fig:pump_assembly}. The miniaturized ion getter pump integrated into the vacuum package is based on a silicon field‑emitter chip developed at IMPT. The main active component is a glass–silicon emitter chip consisting of diced field‑emitter tips and a glass extraction electrode; its fabrication, development, and characterization are described in detail in Refs. \cite{buchta_novel_2023, buchta_glass_2024, buchta_diamond_2025} For the present work, an array of $39\times 39$ diced field emitters was used together with a ring‑shaped electron collector of $3\,$mm diameter, as shown in Fig.~\ref{fig:pump_assembly}a). The electron collector and all pump‑holder components were fabricated at IMPT from $1\,$mm thick fused silica using a LightFab 3D Printer® and a selective laser etching (SLE) process. In this process, the glass is first locally modified by a femtosecond laser and subsequently etched in potassium hydroxide (KOH), such that the modified regions are removed significantly faster than the unmodified bulk material; typical etching times were about $12\,$h.
    
    To realize the electrode function, the holders were metallized using shadow masks with a $50\,$nm Ti adhesion layer followed by $500\,$nm Au, and electrical contact to the external circuitry was provided by pins inserted through through‑glass vias (TGVs) arranged according to the flange design. These pins provide both the electrical connection to the electrodes and mechanical support for the assembly.
    
    For the ion‑collector design, a double‑sided fabrication scheme was employed. On the front side, a cavity of approximately $800\,$µm depth was formed in the fused‑silica holder chip and coated with a $3\,$µm Ti layer to define the ion‑collector electrode. The back side of the chip was first metallized through a shadow mask with $50\,$nm Ti and $500\,$nm Au, and then structured in the LightFab 3D Printer® system to ablate the spiral heater pattern, as shown in Fig.~\ref{fig:pump_assembly}c). The heater resistance was verified via the two low‑voltage copper pins. A remaining glass thickness of about $200\,$µm between the titanium electrode and the heater provides electrical insulation while ensuring sufficient thermal coupling to the Ti getter surface.

    \begin{figure*}
		\centering
		\includegraphics[width = 0.8\textwidth]{"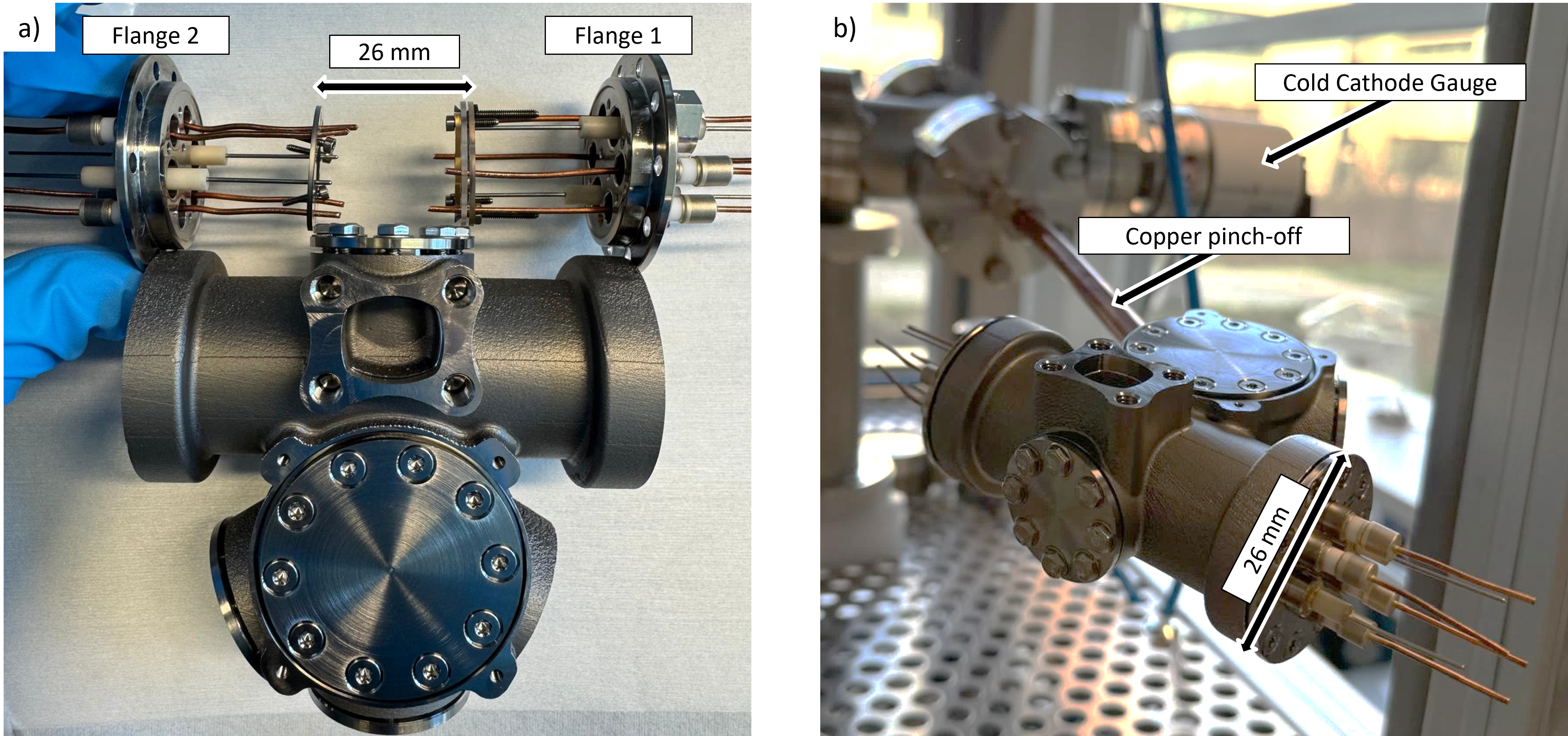"}
		\caption{Integration of the miniaturized ion getter pump into the vacuum chamber. a) Top view of the assembled pump showing pump flange 1 and pump flange 2 inside the additively manufactured titanium chamber. b) Experimental configuration with the chamber connected to the external pumping station via a copper pinch‑off tube and a cold‑cathode gauge mounted close to the pump region for pressure monitoring. The unused ports of the vacuum chamber were sealed with blank flanges. \label{fig:pump_in_chamber}}
	\end{figure*}
    
	\section{Results}
	\subsection{Sr MOT demonstrator}
	We placed a solid block of strontium into the reservoir. The grating was illuminated from the top with a flat-top beam of $2\,$cm diameter that was detuned by $-22\,$MHz from the broad $^1S_0 \rightarrow \,^{1}P_1$ transition at $461\,$nm. The magnetic field gradient was $5.2\,\mathrm{mT}/\mathrm{cm}$. The oven was heated electrically, assisted by an optical heating beam from the top. The base pressure measured by the ion getter pump was $5\times 10^{-9}\,$mbar and increased to $2\times 10^{-8}\,$mbar when the oven was heated.
	
\begin{figure}
		\centering
		\includegraphics[width = 0.45\textwidth]{"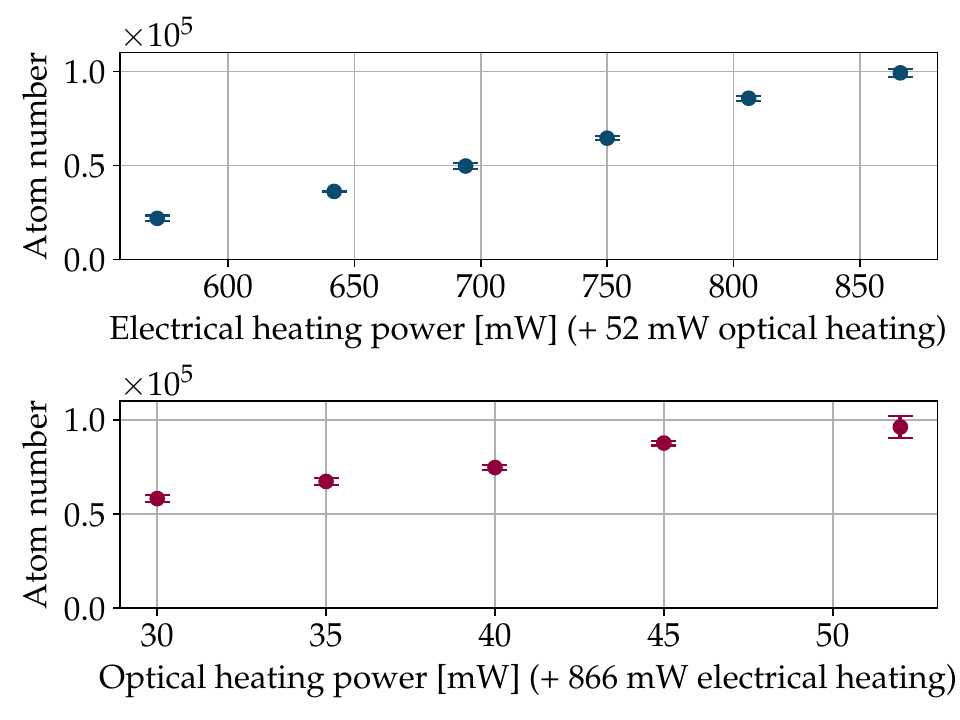"}
		\caption{Trapped atom number in the MOT as a function of the heating power of the oven.\label{fig:atomnumber_vs_heating_power}}
	\end{figure}

	Figure~\ref{fig:atomnumber_vs_heating_power} shows the number of trapped $^{88}$Sr atoms for an optical power of $60\,$mW in the incident MOT beam and varying electrical and optical heating powers. While the optical heating power was scanned, the electrical heating power was set to $855\,$mW. While the electrical power was varied, the optical power was set to $52\,$mW. Compared to our previous results with Yb\cite{pick_low_2025},	a larger heating power is required to evaporate Sr atoms. This can be explained by a reduced thermal contact between the reservoir and the Sr, since no layer of indium was placed in between. Still, the required total heating power of less than $1\,$W is far below that of conventional Sr ovens.
	
\begin{figure}
		\centering
		\includegraphics[width = 0.45\textwidth]{"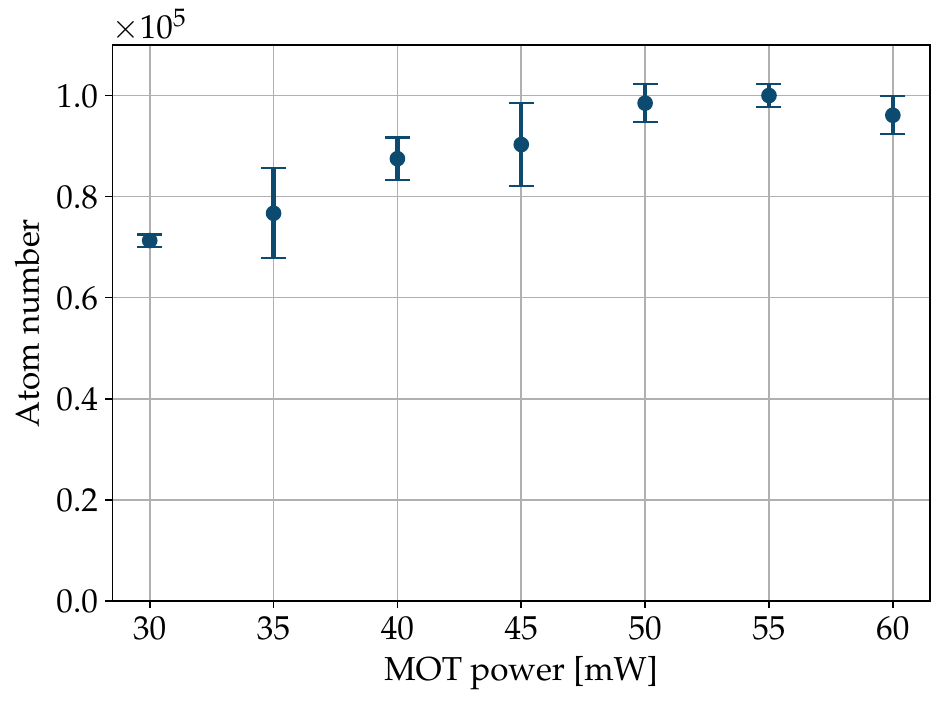"}
		\caption{Trapped atom number in the MOT as a function of the optical power in the MOT laser beam.\label{fig:atomnumber_vs_MOT_power}}
	\end{figure}

	Figure~\ref{fig:atomnumber_vs_MOT_power} shows the atom number in the MOT as a function of optical power in the MOT beam. The oven was heated with $855\,$mW electrical power and $52\,$mW optical power. The atom number increases with increasing MOT power, and settles around $55\,$mW. The trapped atom number $N=\dot{N}\cdot\tau$ is determined by the product of the MOT loading rate $\dot{N}$ and the lifetime $\tau$. The comparably low atom number suggests a relatively short liftetime. Illumination of the MOT with lasers resonant to the repumping transitions $^3P_0 \rightarrow \,^{3}S_1$ and $^3P_2 \rightarrow \,^{3}S_1$ at $679\,$nm and $707\,$nm did not have an observable effect on the atom number. This suggests that the MOT lifetime was not limited by optical losses, and that losses due to collisions with background atoms dominated. At a pressure of $2\times10^{-8}\,$mbar, the expected vacuum-limited MOT lifetime is $150\,$ms \cite{nagel_magnetic_2003}. For the measured atom numbers, this corresponds to MOT loading rates on the order of $10^6\,\mathrm{atoms}/\mathrm{s}$. However, the vacuum pressure has been measured at the ion getter pump outside of the vacuum chamber. The MOT is located directly above the hot oven, so that a higher background pressure can be expected at this position. This means that the actual MOT loading rates might be higher than this estimate.
	
	The trapped atom number can be increased by reducing the vacuum pressure. This might be achieved by improving the thermal contact between the Sr and the reservoir, resulting in a lower temperature of the reservoir that is required to evaporate atoms. In our previous results with Yb \cite{pick_low_2025}, the thermal contact was significantly better, resulting in a lower vacuum pressure. This was accomplished by placing a thin layer of indium in between Yb and the reservoir. Another option would be to fill the reservoir by evaporation of atoms from an external source and subsequent deposition on the glass substrate\cite{manginell_in_2012, schwindt_highly_2016}.
	Although the trapped atom number can be further increased, the results presented here show that the demonstrated setup is a very promising approach for the realization of highly miniaturized quantum sensors.
	
	\subsection{Vacuum pump}
	The pump assembly was integrated into a separate but identical vacuum chamber. 
	The silicon field‑emitter chip and the titanium ion‑collector assembly were mounted on the two pump flanges, as shown in Figs.~\ref{fig:pump_assembly} and \ref{fig:pump_in_chamber}. The ring‑shaped electron collector and the ion‑collector electrode were separated by approximately $26\,$mm, limited by the length of the high‑voltage feedthrough pins inside the chamber (see Fig.~\ref{fig:pump_in_chamber}a). The assembly was evacuated by an external pumping station, connected to the chamber via a copper pinch‑off tube. A cold‑cathode gauge, shown in Fig.~\ref{fig:pump_in_chamber}b), monitored the pressure during the experiments. During all measurements, the chamber stayed attached to the copper pinch‑off flange. Both the high‑voltage and heater connections were enclosed in 3D‑printed housings for insulation and mechanical protection.
	
	A characterization of a stand-alone demonstrator of the magnet-free ion getter pump, including a long‑term pumping test in a dedicated UHV chamber, is reported in Ref.~\cite{buchta_verbundprojekt_2025}. In the following, we focus on the integration of the pump into the additively manufactured vacuum chamber.
	
    Although residual leaks and imperfect sealing of the current assembly prevented a quantitative assessment of the intrinsic pump performance, the measurements provide indications of heater and pump operation accompanied with a pumping effect. Due to the leaks in the test setup, the pumping effect was observed in the $10^{-4}\,$mbar regime only. With the gate to the external pumping station closed, the heater and the ion pump noticeably affected the equilibrium pressure in the vacuum chamber. This observation suggests that the pump provides a positive pumping effect, although no demonstration in the UHV was possible in this setup. The overall behavior is dominated by the leak-limited gas load and possible additional gas desorption and scrubbing at high emission currents. Therefore, this experiment is best regarded as an initial integration study demonstrating assembly compatibility and basic functional operation of the pump and heater, rather than the ultimate performance of the pump concept. This proof-of-concept makes the magnet-free ion getter pump a valuable component of a future fully-integrated compact MOT setup for mobile and commercial quantum sensors.

	\section{Conclusion}
	We have demonstrated a low-SWaP source of laser-cooled Sr atoms. The high degree of miniaturization is achieved by combining different innovative technologies, namely a single-beam MOT configuration, a chip-based atomic oven, an additively manufactured titanium vacuum chamber, and custom-sized vacuum flanges. Generating the MOT with a grating MOT chip makes full use of the compact, chip-like geometry of the glass-based atomic oven. Although the vacuum chamber could have been manufactured by conventional machining methods alone, the demonstrated additive manufacturing of a titanium vacuum chamber opens the path towards more complex chamber designs based on titanium. The trapped atom number can be increased by improving on the vacuum pressure, for example by improving the thermal contact between the Sr and the oven reservoir. A further reduction of the temperature of the trapped atoms can be achieved by implementing a MOT operating on the second cooling transition at $689\,$nm, which is also supported by the grating MOT chip \cite{bondza_two_2022}. We have also developed a miniaturized ion getter pump that does not require a magnetic field and that can be inserted into the vacuum chamber. Although the characterization of the pump was limited by a leak in the test setup, compatibility with the additively manufactured vacuum chamber and a pumping effect were demonstrated.
	
	The presented technologies enabled the realization of a miniaturized vacuum package for magneto-optical trapping of Sr, thereby paving the way towards next-generation, highly compact quantum sensors.

	\begin{acknowledgments}
	We acknowledge funding from the joint project ``Innovative Vacuum Technology for Quantum Sensors'' (InnoVaQ) funded by the German Federal Ministry of Research, Technology and Space (BMFTR) as part of the funding program ``quantum technologies – from basic research to market''. (Contract numbers: 13N15915, 13N15916, 13N15917, 13N15918 and 13N15919). The project has also received funding from the State of Lower Saxony, Hannover, Germany,  through Niedersächsisches Vorab. We further acknowledge funding by the Deutsche Forschungsgemeinschaft (DFG, German Research Foundation) under Germany’s Excellence Strategy – EXC-2123 QuantumFrontiers – 390837967. 
	 
	\end{acknowledgments}

%\section*{Data Availability Statement}

\bibliography{bibliography}
		
\end{document}